\documentclass[prl,preprint,longbibliography]{revtex4-1}%
\usepackage{amsmath}
\usepackage{amsfonts}
\usepackage{amssymb}
\usepackage{graphicx}%
\usepackage{color}
\providecommand{\U}[1]{\protect\rule{.1in}{.1in}}
\begin{document} \title{Bursting dynamics of viscous film without circular symmetry: the effect of confinement} \author{Mayuko Murano and Ko Okumura\footnote{okumura@phys.ocha.ac.jp}} \affiliation{Department of Physics and Soft Matter Center, Ochanomizu University, 2--1--1, Otsuka, Bunkyo-ku, Tokyo 112-8610, Japan.}

\begin{abstract}
We experimentally investigate the bursting dynamics of confined liquid film suspended in air and find a viscous dynamics distinctly different from the non-confined counterpart, due to lack of circular symmetry in the shape of expanding hole: the novel confined-viscous bursting proceeds at a constant speed and a rim formed at the bursting tip does not grow. 
We find a confined-viscous to confined-inertial crossover, as well as a nonconfined-inertial to confined-inertial crossover, at which bursting speed does not change although the circular symmetry in the hole shape breaks dynamically.
\end{abstract}\maketitle

\flushbottom
\thispagestyle{empty}

{\it Introduction.\textemdash} 
Bursting of liquid thin film is important in many fields and plays crucial roles in many industrial processes: it is observed with polymer foams, in glass furnaces, and during volcanic eruption. This phenomenon is familiar to everyone as the rupture of soap films, i.e., liquid films suspended in air \cite{1959Taylor,1969Mysels,1990Davidson,2007Stannarius}. 
Different dynamics have been revealed for bursting of fluid film surrounded by viscous liquid through studies on antibubbles \cite{2012Etoh, 2013Fu} and emulsions etc. \cite{2006Etienne, 2010Eri}
Dewetting, i.e., bursting of a film deposited on a substrate, is another topic that has attracted much attention \cite{1991Brochard,2005Raphael,2007Raphael}. 
More practically, bursting is an important tool for generation of droplets \cite{1990Davidson,1998Spiel,2009Villermaux,2010Stone,2015Bird,2016Wu,2015Ligoure} or emulsions \cite{2014Stone} with controlled sizes because at a later stage a liquid rim formed at the bursting tip destabilizes to break into droplets. The size of droplets is a crucial factor in various fields such as environmental science \cite{1981Wu} or disease transfer \cite{2005Mark} because contamination results often via droplets.
Bubble bursting is also important to bioreactor efficiency, because bursting is one of the causes of cell damage \cite{1992Hulle,2011Hu,2014Liu}.
 
Taking a closer look at the bursting of liquid film suspended in air, which is the base of all the above mentioned issues, only two scaling regimes have been established, one is called inertial regime and the other viscous (or viscoelastic) regime.
For almost inviscid films such as soap films or smectic films, the so-called inertial regime has been reported, in which
a film bursts at a constant velocity \cite{1959Taylor,1969Mysels,1990Davidson,2007Stannarius}. 
This velocity is known as the Taylor-Culick velocity
$U_{C}=\sqrt{2\gamma/(\rho h)}$, which results from the balance between capillarity and inertia ($\gamma$, $\rho$ and $h$ are the surface tension, density of the liquid, and the thickness of the film, respectively).
In contrast, for ultra viscous films of polymer solutions, the viscous regime has been reported, in which bursting accelerates exponentially with time \cite{1995Debregeas,1998Debregeas}.  
The characteristic time for the exponential dynamics is given by $\tau_{v}\sim\eta h/\gamma$ with $\eta$ the viscosity of liquid. The physical origin of the viscous regime has been a contentious issue. 
Some considered that the  regime is realized by elastic feature of entangled polymers and termed the regime ``viscoelastic" \cite{1995Debregeas,1998Debregeas,Capillaritytext}, while others showed that this regime can be reproduced in a purely viscous liquid via numerical simulation and analytical theory \cite{1999Gueyffier,2002Durst,2009Bush}.
 
While there still remains a controversy even for nonconfined three-dimensional (3D) bursting, studies have been very limited for bursting of spatially confined films suspended in air despite their fundamental importance. 
In fact, a recent study on liquid-drop coalescence \cite{2010Eri,Okumura} suggests the importance of the study of confined quasi two-dimentional (2D) bursting in air, demonstrating distinct confinement effects on the dynamics.
In addition, such a confined dynamics should play fundamentally important roles in many practical applications such as related to foams \cite{2007Langevin,2011Anne-Laure,foam}, given that any films in foam are inherently confined by Plateau borders. 
Here, we study the dynamics of viscous film bursting in a confined geometry of the Hele-Shaw cell. 
As a result, we show by clear agreement between theory and experiment that bursting of the film sandwiched by the walls exhibits a remarkably different viscous dynamics compared with the nonconfined case in 3D.
In the nonconfined case, the bursting accelerates with time without rim formation in the viscous regime (nonconfined-viscous regime), and the bursting proceeds with a constant velocity with a growing rim in the inertial regime (nonconfined-inertial regime). 
However, in the confined case in quasi 2D using the Hele-Shaw geometry, 
films burst at a constant velocity with a rim of fixed size in the viscous regime (confined-viscous regime).
As for the confined-inertial dynamics, the bursting proceeds with a constant velocity, which is equal to the nonconfined-inertial velocity.
We remarkably confirmed directly the nonconfined-to-confined inertial crossover with no change in bursting velocity when the circular symmetry of the shape of the expanding hole is dynamically broken.

\begin{figure}[h]
\includegraphics[width=0.85\linewidth]{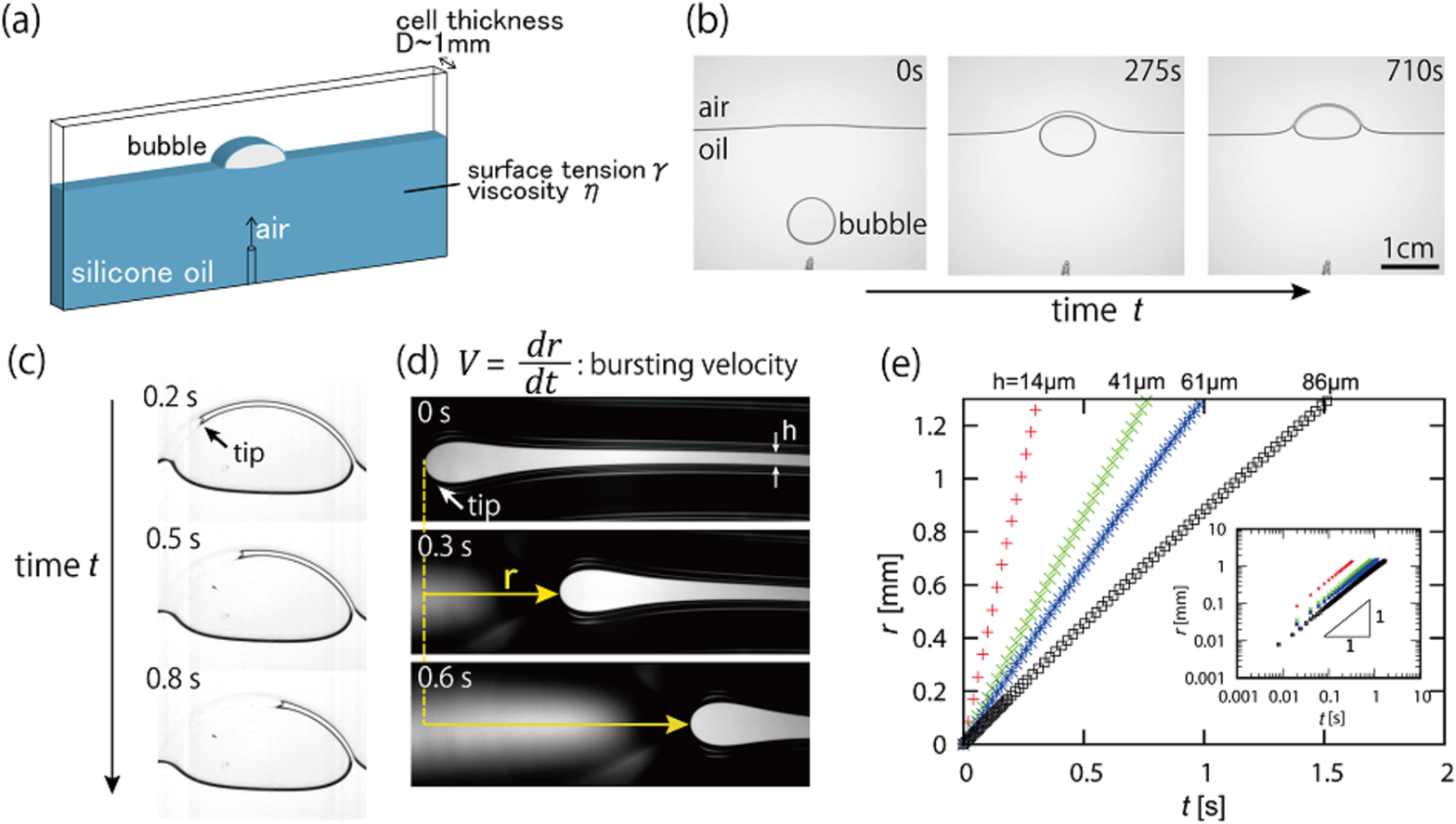}
\caption{ (a)~Experimental setup. (b)~Dynamics of a bubble in the cell well
before the film bursting starts. (c)~Image sequence of a bubble bursting. (d)~Magnified snapshots of a
bursting tip taken at a fixed time interval. The tip position $r$ is
measured from the position in a certain reference snapshot. (e)~Bursting tip
position $r$ vs. time $t$ obtained from films with different thickness $h$ ($D=$ 0.97 mm, $\eta=$ 9.7 Pa$\cdot$s). The
shapes of tips in (c) and (d) look different because focus points of photograph are different.\label{Fig1}}
\end{figure}

{\it Results. (a) Experimental\textemdash} 
We fabricate a Hele-Shaw cell from two
acrylic plates separated by the distance $D$ (Fig.~\ref{Fig1}(a)). 
The cell thickness $D$ is adjusted by spacers, and is directly measured using the laser distance sensor
(ZS-HLDS5, Omron). We fill the
cell with silicon oil ($\eta=$ 0.97 $\sim$ 48 Pa$\cdot$s, $\rho=$
965 kg/m$^{3}$, $\gamma=$ 20 mN/m) and inject a bubble from
the bottom of the cell. The bubble slowly rises in the viscous oil up to the
liquid-air interface (Fig.~\ref{Fig1}(b)), stays there for a while, and
suddenly bursts and disappears. The bursting is recorded by a high
speed camera (FASTCAM SA-X, Photoron) with a macro lens (AF-S Micro NIKKOR 60mm f/2.8G ED, Nikon) or a microscope lens (PLN10$\times$ or PLN4$\times$, OLYMPUS). The point where bursting starts is not
controlled and the point can be any places (e.g. places near the edges) of
the liquid film encapsulating the bubble.  
We wait till the film bursts without any artificial trigger, which means that the thickness at the moment of bursting is not controlled because the film of the bubble keeps thinning while the bubble stays at the liquid-air interface \cite{2007Eri}. 
Because of this, when the bursting starts shortly after the bubble reaches the interface, the thickness of the bursting film is relatively thick. In such a case, the bursting is found to be in a viscous regime (as shown in (b) below).
On the contrary, when the bursting starts after the film thickness becomes significantly thin, the dynamics is found to be in the inertial regime as discussed in (c) below. 

{\it (b) Bursting dynamics in the viscous regime\textemdash} 
Figure~\ref{Fig1}(c) though (e)
demonstrate experimental results on the dynamics of bursting tips in the viscous regime (see Supplemental Material (SM) movie 1 \cite{movie}). Snapshots
of a whole bubble and a magnified tip taken after bursting starts from the
left edge are shown in Fig.~\ref{Fig1}(c) and (d), respectively. 
It is demonstrated by the three
snapshots taken at a regular interval in Fig.~\ref{Fig1}(d) and by the relation
between the tip position $r$ and elapsed time $t$ given in Fig.~\ref{Fig1}(e)
that the bursting velocity is constant, while the velocity depends
on the film thickness as shown in Fig.~\ref{Fig1}(e). 
Although a rim exists at the tip, the rim does not grow as the bursting proceeds, as indicated in
Fig.~\ref{Fig1}(d) and verified in Sec.~A of SM \cite{SM}.

As shown in Fig.~2(a), we add particles 
(techpolymer MBX-20, SEKISUI PLASTICS) into oil to see the flow inside the film 
(the number of the added particles are made small to minimize the disturbance of the flow by them). 
The movement of particles implies that the $y$ and $z$ components of the flow are practically zero (see movie 2 \cite{movie} and Sec.~B of SM \cite{SM}) and that the remaining $x$ component is constant in the $z$ direction (see Sec.~C of SM \cite{SM}). 
In addition, the length of the disturbed region $L$ increases with $D$, which is justified in the following manner.
In Fig.~2(b), the $x$ component $v$ of flow velocity normalized by bursting velocity $V$ is given as a function of the distance from the tip $x$ on the basis of the particle-tracking analysis. 
As revealed in the inset of Fig.~2(b), the flow profile $v$ exhibits an exponential decay with $x$, from which we determine the decay length $L$ (See Sec.~D of SM \cite{SM} for further details). 
Figure~2(c) shows that $L$ increases with $D$. 

As for the shape of a bursting tip, our analysis indicates that the radius of curvature at the bursting tip $R$ is determined by $h$.
Figure~2(d) shows that $R$, which is determined by fitting the shape of the rim to a parabolic function, as a function of the film thickness $h$ for different $D$ and $\eta$. 
This indicates that $R$ increases with $h$ and is independent of $D$ or $\eta$.

\begin{figure}[h]
\includegraphics[width=1\linewidth]{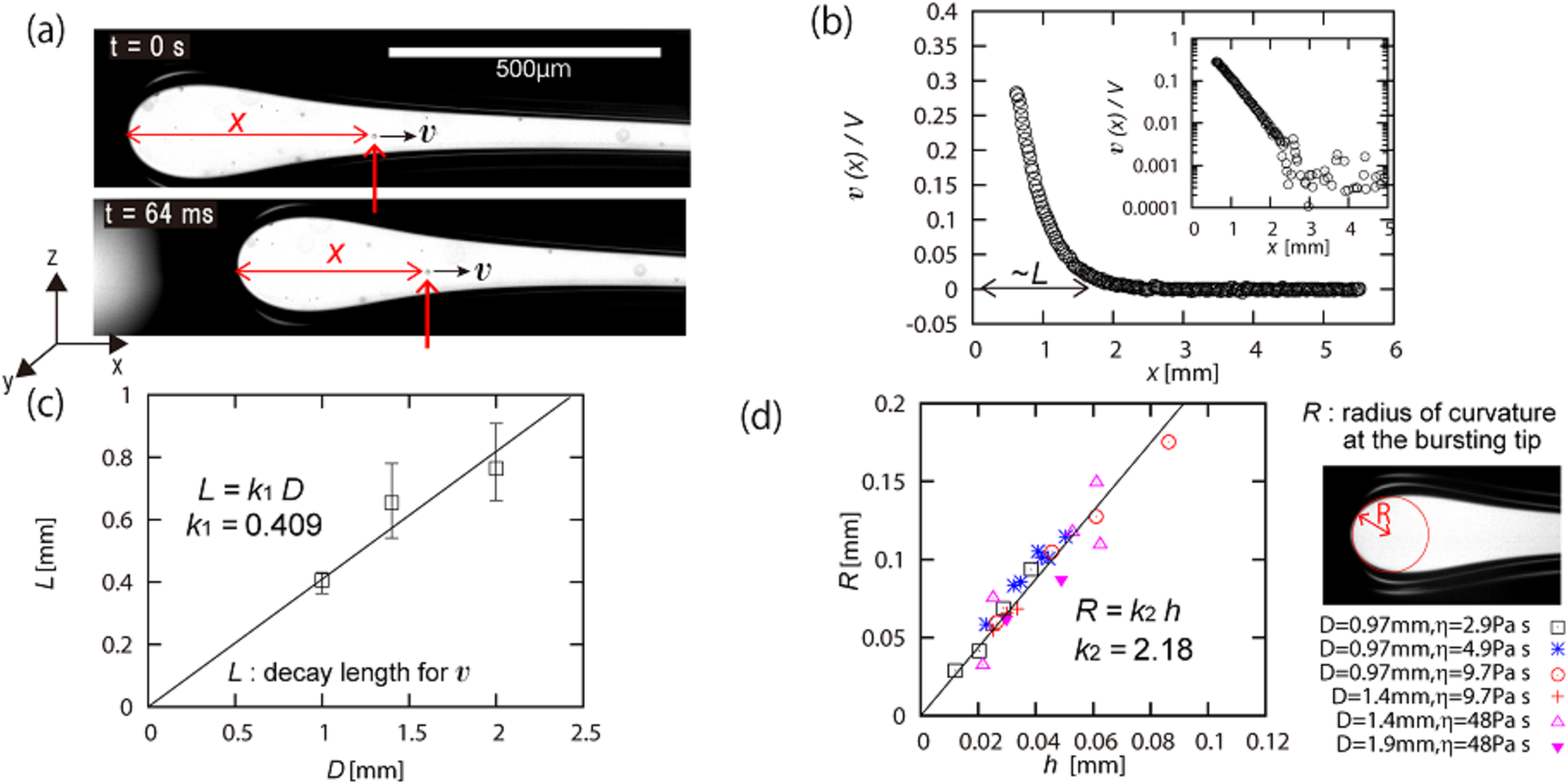} 
\caption{(a)~Example images for analysis of
the flow inside the film by adding a small number of particles to oil ($D=$1.0 mm, $\eta =$ 4.8 Pa$\cdot$s,  $h=$ 44 $\mu$m). 
The thick vertical arrows indicate the horizontal position of the particle in question.
(b)~The velocity $v$ normalized by the bursting velocity $V$ vs. distance from the tip $x$ ($D=$1.0 mm, $\eta =$ 4.8 Pa$\cdot$s,  $h=$ 44 $\mu$m), obtained by the velocity change of a single particle.
inset: semilogarithmic plot of the same data showing an exponential decay of $v$.
(c)~Decay length $L$ vs. cell thickness $D$. Each point shows the average of four data and error bars indicate maximum and minimum values. Straight line shows $L=k_1 D$ with $k_1 =0.409 \pm 0.026$.
(d) Radius of curvature at the bursting tip $R$ vs. thickness $h$ on a log-log scale. 
The straight line shows $R=k_2 h$ with $k_2 =2.18 \pm 0.05$.}
\label{Fig2}%
\end{figure}

Theoretically, the observed bursting velocity can be explained by a global balance between the surface and dissipative energies under an assumption consistent with Fig.~2(c). 
The bursting is driven by decrease in surface energy of the film, which is dimensionally estimated as $d(\gamma Dr)/dt = \gamma DV$ per unit time. 
The viscous dissipation is estimated as follows. 
As discussed above, only nonnegligible component of the velocity vector is the $x$ component ($\sim V$). 
The velocity gradient for this component inside the film develops in both $x$ and $y$ directions and both gradients scale as $V/D$. 
Note that in the $y$ direction, the film is sandwiched by walls and thus the flow profile is a Poiseuille type with the velocity gradient $V/D$,
and the velocity gradient in the $x$ direction $V/L$ scales as $V/D$ under the assumption $L \sim D$, which is consistent with Fig.~2(c).
The velocity gradient $V/D$ originating from these two origins is developed in a volume of the order of $hDL$ with $L \sim D$. 
Thus, the viscous energy dissipation is described as $\eta (V/D)^2 hD^2$. 
From the balance of the two energies, $\gamma DV \sim \eta (V/D)^2hD^2$ , we obtain the bursting
velocity $V = U_{\eta}$ where
\begin{equation}
U_{\eta}=k\frac{\gamma}{\eta}\frac{D}{h} \label{V}%
\end{equation}
with a numerical coefficient $k$ to be determined experimentally.
(See Sec.~E of SM \cite{SM} for another derivation of Eq.~(1), which is consistent with Fig.~\ref{Fig2}(d).)

As shown in Fig.~3, Eq.~(1) agrees well with experimental results. The results of measurements of $V$ as a function of $h$ for various $D$ and $\eta$ are shown in Fig.~3(a). The same data are replotted with both axes renormalized according to Eq.~(1) in Fig.~3(b), in which a clear collapse of the data is demonstrated with $k$ in Eq.~(1): $k = 0.0350 \pm 0.0005$.

\begin{figure}[h]
\includegraphics[width=0.5\linewidth]{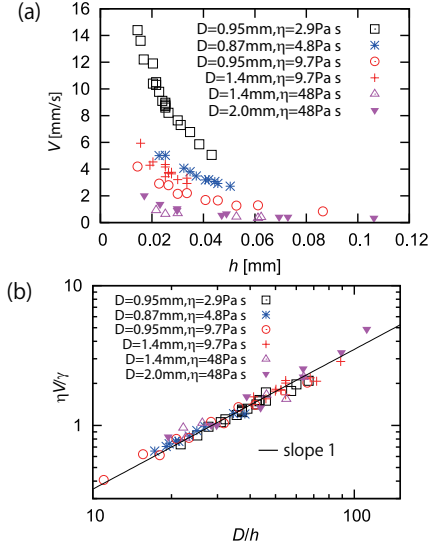} \caption{(a)~Bursting
velocity $V$ vs. film thickness $h$. (b)~Normalized velocity
$\eta V/\gamma$ vs. normalized thickness $D/h$.\label{Fig3}}
\end{figure}

{\it (c) Bursting dynamics in the inertial regime\textemdash} 
As the Reynolds number Re increases, the confined-viscous regime exhibits a crossover to the confined-inertial regime at Re $\sim 1$ 
(Re $\sim 10^{-3}$ in Fig.~\ref{Fig1} $\sim$ to \ref{Fig3}).
Here, Re is estimated as $\rho VD/\eta$ by considering the ratio of the inertial force $F_i=d(MV)/dt=\rho DhV^2$ (with the mass of a rim $M=\rho rDh$) to the viscous force $F_v \sim \eta V hD^2/D^2$ (see Sec.~F of SM for another derivation of the expression for Re, which supports $L \sim D$ suggested in Fig.~2(c) \cite{SM}). 
Note that the bursting in the confined-inertial regime is assumed to proceed at a constant speed ($V=dr/dt$) as experimentally observed. 

The bursting velocity $U_i$ in the confined-inertial regime is derived from the balance between capillary force $2\gamma D$ and inertial force $F_i$, which results in $U_{i}=\sqrt{2\gamma/(\rho h)}$.
This velocity is the same with that in the nonconfined 3D case including the numerical coefficient, i.e., $U_i=U_C$.

\begin{figure}[ptb]
\includegraphics[width=1\linewidth]{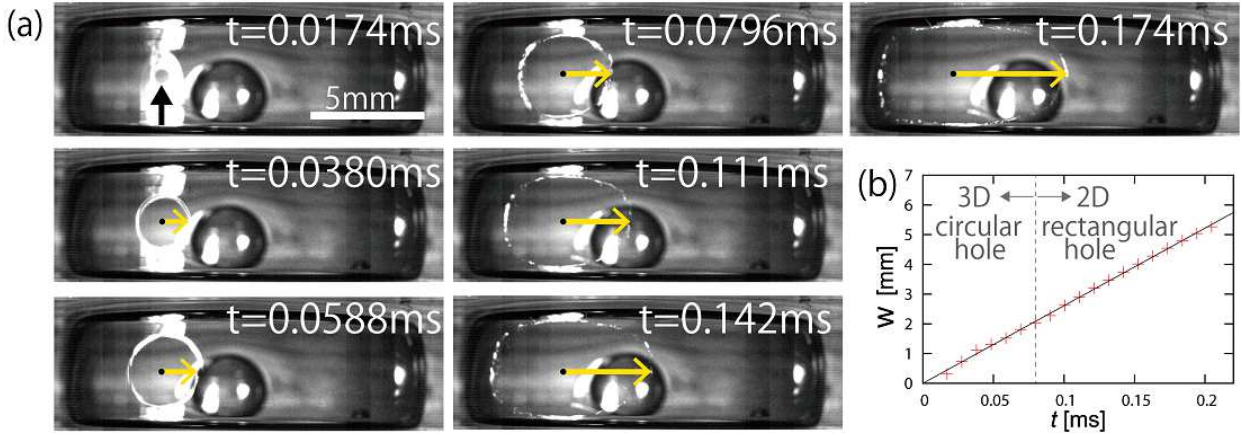} \caption{(a)~Sequential images
of a bursting bubble taken from above ($D=$ 6 mm, $\eta$ =0.0965 Pa s). 
The vertical arrows in the first shot indicate the nucleation point of bursting.
The horizontal yellow arrows define the half width of the hole $w$. 
The small bubble observed at the center of each snapshot is irrelevant to the
bursting in question. (b)~Half width of the hole $w$ (indicated by horizontal arrows in snapshots) vs. time
$t$. The constant velocity ($\simeq$26.1 m/s) is maintained during the change in the hole shape
from circular to rectangular ones at $t=$ 0.0729 ms, indicated by the dashed line. This dimensional crossover occurs before $w$ becomes $D/2$, because films exist on the side walls; the hole size in the direction of cell thickness never reaches $D/2$. \label{Fig4}}
\end{figure}

The confined-inertial bursting predicted above can be observed in experiment for Re $\gg$ 1 (see Fig.~\ref{Fig4}, in which Re $\sim 20$). 
In Fig.~\ref{Fig4}(a), the expanding hole grows at first with maintaining the circular shape as in the case of the nonconfined 3D film and then changes its shape to the quasi-rectangular one, but the bursting speed is unchanged throughout bursting as shown in Fig.~\ref{Fig4}(b).
This observation is consistent with the above prediction in the following manner.
First, the observed initial 3D bursting is in the inertial regime because, if a bursting proceeds at a constant speed for a 3D film, the bursting velocity is $U_C$, as we mentioned in the second introductory paragraph. 
Second, since Re $\gg 1$ is satisfied, the observed rectangular bursting is expected to be in the confined-inertial regime and thus proceeds with the velocity $U_i$, which is observed to be equal to the nonconfined-inertial velocity $U_{C}$, as predicted.
Finally, the rim growth, which we suppose in the deviation of $U_i$, is consistent with experimental observation. 
This is because, at the bursting tip drop generation can be seen, which is caused by fragmentation of an amply growned rim into small droplets (see SM movie 3 \cite{movie}).

{\it Discussion.\textemdash} 
At very short times after the nucleation of an initial hole for bursting before the confined-viscous regime sets in, the film seems to rupture at a velocity dozens times higher than $U_{\eta}$. 
If we could capture this initial stage of bursting, we would see the change in the shape of the
expanding hole from circular (nonconfined 3D) to quasi-rectangular (confined quasi-2D) in the viscous regime as
observed in the
inertial bursting. However,
this ultrafast regime is difficult to capture and requires a separate study. This is because the control of the point where bursting starts is technically difficult and the time scales for the ultrafast initial regime and the viscous regime are extremely different.

In Ref. \cite{2007Kliakhandler}, the authors confined a film between two needles and punctured the film by another needle to measure the bursting velocity and reported a bursting velocity different from ours. This difference may originate from difference between the needle and Hele-Shaw geometries and/or significant difference in characteristic length scales of the two experiments.

The present work could make a significant contribution to the field of bursting film in air, given that only a few scaling regimes have been known despite the long history of research. For example, this work provides fundamentally important knowledge for understanding the dynamics of foams in general \cite{foam}. 
Controlling the rim growth with the aid of Eq.~(\ref{V})  may also be useful for environmental problems or industrial applications associated with generation of droplets \cite{1981Wu,2005Mark,2016Gupta}. Furthermore, Eq.~(\ref{V}) and the remarkable invariance of the bursting speed ($U_i=U_C$) at the 3D to quasi-2D crossover in the inertial regime could be useful for measuring the thickness of confined liquid film in a wide range from micron- to nanometer-scales.

\section{Acknowledgements}
The authors thank Natsuki Kimoto for making trials for the initiation of this research project. This research was partly supported by Grant-in-Aid for Scientific
Research (A) (No. 24244066) of JSPS, Japan, and by ImPACT Program of Council
for Science, Technology and Innovation (Cabinet Office, Government of Japan).
M. M. is supported by the Japan Society for the Promotion of Science Research
Fellowships for Young Scientists (No. 16J00871).

%

\clearpage

\end{document}